\documentclass{PoS}
\usepackage{amsmath, empheq, amsfonts, amssymb}
\usepackage{subfig}
\title{Study of the $DK K$ and $DK \bar{K}$ systems}

\ShortTitle{Study of the $DK K$ and $DK \bar{K}$ systems}

\author{\speaker{V.~R.~Debastiani}\\
       Departamento de F\'{\i}sica Te\'orica and IFIC, Centro Mixto Universidad de Valencia-CSIC Institutos de Investigaci\'on de Paterna, Aptdo. 22085, 46071
        Valencia, Spain\\
       E-mail: \email{vinicius.rodrigues@ific.uv.es}}

\author{J.~M.~Dias\\
        Departamento de
F\'{\i}sica Te\'orica and IFIC, Centro Mixto Universidad de
Valencia-CSIC Institutos de Investigaci\'on de Paterna, Aptdo.
22085, 46071 Valencia, Spain\\
Instituto de F\'{\i}sica, Universidade de S\~{a}o Paulo,
C.P. 66318, 05389-970 S\~{a}o Paulo, SP, Brazil\\
        E-mail: \email{jorgivan.morais@ific.uv.es}}

\author{E.~Oset\\
        Departamento de
F\'{\i}sica Te\'orica and IFIC, Centro Mixto Universidad de
Valencia-CSIC Institutos de Investigaci\'on de Paterna, Aptdo.
22085, 46071 Valencia, Spain\\
        E-mail: \email{oset@ific.uv.es}}

\abstract{Using the Fixed Center Approximation to Faddeev equations we have investigated the $DKK$ and $DK\bar{K}$ three-body systems, considering that the $D^*_{s0}(2317)$ acts as the heavy cluster in both cases, generated from the $DK$ interaction in isospin 0. For the $DK\bar{K}$ system we have found evidence of a state with $I(J^P)=1/2(0^-)$ and mass about $2833 - 2858$ MeV, above the threshold of  $D^*_{s0}(2317)\bar{K}$. Our results indicate that this state is dominated by a $Df_0(980)$ component, then it could be searched for in the $\pi \pi D$ invariant mass. On the other hand, no clear evidence related to a state from the $DKK$ interaction is found.}

\FullConference{XVII International Conference on Hadron Spectroscopy and Structure - Hadron2017\\
		25-29 September, 2017\\
		University of Salamanca, Salamanca, Spain}

\begin{document}

\section{Introduction}

The study of three-body systems is one of the starting points in the study
of nuclei and nuclear dynamics. The traditional Quantum Mechanical approach
to this problem is based on the Faddeev equations. The simplicity of the Faddeev
equations is deceiving since in practice its evaluation is very involved and one
approximation or another is done to solve them. One popular choice is the use
of separable potentials to construct the two-body scattering amplitudes via the
Alt-Grassberger-Sandhas (AGS) form of the Faddeev equations. Another way to tackle
these three-body systems is using a variational method.

In our work \cite{DKK} we use the Fixed Center Approximation (FCA) to study systems of three mesons: the $DKK$ and $DK\bar{K}$. The main idea behind the FCA is to break the three-body problem in terms of one heavy cluster, generated by the interaction of two components, while a third particle (lighter than the cluster) undergoes multiple scattering with the components of the cluster. In both systems we start the problem with the $D_{s0}^*(2317)$ molecule, formed from the $DK$ interaction in isospin 0. This interaction is well studied in works using chiral Lagrangians and unitary approach \cite{Gamermann}, and is also supported by analysis of lattice QCD data \cite{lattice}. On top of the $D_{s0}^*(2317)$ cluster, another $K$ (or $\bar K$) is introduced, which will interact with the $D$ and $K$ components of the molecule.

\section{Formalism}

The FCA mechanism is illustrated in Fig.~\ref{FCA}, representing equations \eqref{Faddeev}:
\begin{figure}
  \centering
  \includegraphics[width=0.95\textwidth]{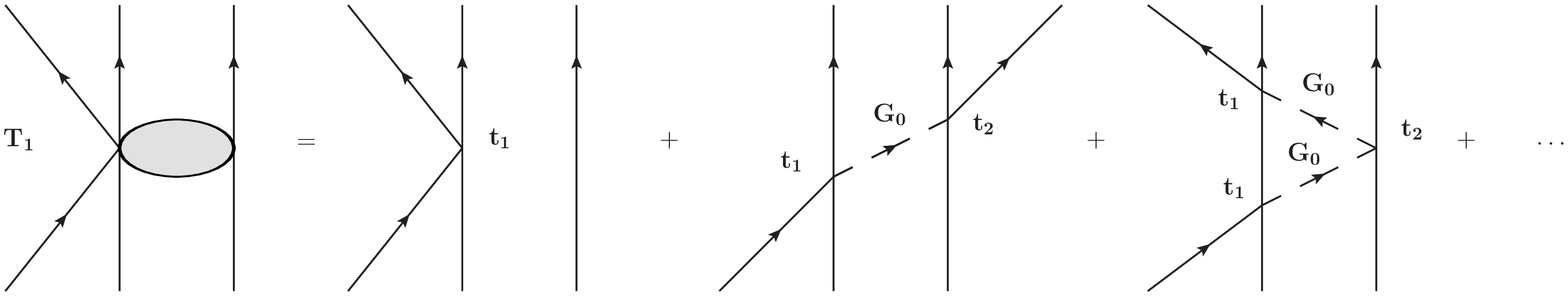}\\
  \caption{Diagrams of Fixed Center Approximation to Faddeev equations.}\label{FCA}
\end{figure}
\begin{align}\label{Faddeev}
\begin{aligned}
  &T_1 = t_1 + t_1 \,G_0 \,t_2 + t_1 \,G_0 \,t_2 \,G_0 \,t_1 + ...
  \qquad\qquad &\Rightarrow T_1 = t_1 + t_1 \,G_0 \, T_2
  \\
  &T_2 = t_2 + t_2 \,G_0 \,t_1 + t_2 \,G_0 \,t_1 \,G_0 \,t_2 + ...
  \qquad\qquad &\Rightarrow T_2 = t_2 + t_2 \,G_0 \, T_1
  \\
\end{aligned}
\end{align}
where $t_1$ and $t_2$ are the two-body amplitudes of the external particles with the left or right component of the cluster, respectively, and $G_0$ is the propagator of the external particle inside the cluster. $T_1$ describes the multiple scattering starting from the left while $T_2$ the one from the right, and the total three-body amplitude is given by the sum $T = T_1 +T_2$.\\

\begin{figure}[h!]\centering
\includegraphics[width=\textwidth]{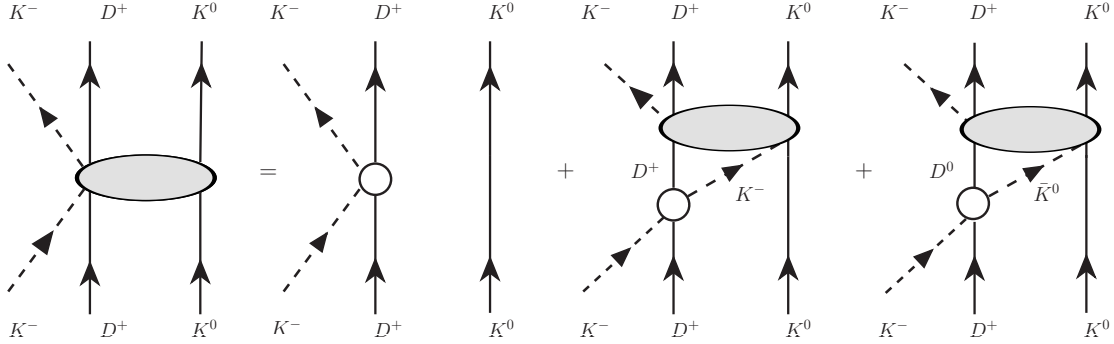}
\caption{Diagrams for the $K^-$ multiple scattering of the process $K^-[D^+K^0] \to K^-[D^+K^0]$. The white circles indicate two-body amplitudes of $D\bar{K} \to D\bar{K}$, while the gray bubbles are associated with three-body amplitudes of $DK\bar{K}$.}\label{diag1}
\end{figure}

In the following we describe the $DK\bar{K}$ system, since the $DKK$ is analogous and details can be found in Ref.~\cite{DKK}. In charge basis we have three channels represented by $T_1$ in Fig.~\ref{FCA}:  (1)~$K^-[D^+K^0]$, (2)~$K^-[D^0K^+]$ and (3)~$\bar{K}^0[D^0K^0]$, when the $\bar{K}$ interacts with the $D$ component of the cluster $[DK]$; and represented by $T_2$ we have (4)~$[D^+K^0]K^-$, (5)~$[D^0K^+]K^-$ and (6)~$[D^0K^0]\bar{K}^0$, when the $\bar{K}$ interacts with the $K$ component. The channels (3) and (6) are used as intermediate charge-exchange steps, an extra feature adopted in our work, as exemplified in the last diagram of Fig.~\ref{diag1}. Using this notation we have a $6\times6$ system of equations. In Fig.~\ref{diag1} we show one example of the diagrams considered in the $(1) \,\to\, (1)$ scattering, which is given by Eq.~\eqref{t11}.
\vspace{-10pt}
\begin{eqnarray}\label{t11}
T^{\rm FCA}_{11}(s)=t_1+t_1\,G_0\,T^{\rm FCA}_{41}+t_2\,G_0\,T^{\rm FCA}_{61}\, ,
\end{eqnarray}
where $t_1$ and $t_2$ are the $D^+K^-\to D^+K^-$
and $D^+K^-\to D^0\bar{K}^0$ two-body scattering amplitudes, respectively.
In this way we can systematically write the three-body amplitudes in terms of the two-body amplitudes and the propagator of $\bar{K}$ inside the cluster:
\begin{eqnarray}
\label{multsca}
T^{\rm FCA}_{ij}(s)=V^{\rm FCA}_{ij}(s)+\sum\limits_{l=1}^{6}
 \tilde{V}^{\rm FCA}_{il}(s)\,G_0(s)\,T^{\rm FCA}_{lj}(s)\, ,
\end{eqnarray}
where $V_{ij}$ and $\tilde{V}_{il}$ are the elements of the matrices describing all the transitions between the 6 channels, written in terms of two-body amplitudes of $DK \to DK$ and $K\bar{K} \to K\bar{K}$ from Refs.~\cite{Gamermann,Oller}. The $V_{ij}$ matrix contains the elements of single-scattering, while $\tilde{V}_{il}$ contains the cases of double scattering, when $\bar{K}$ propagates inside the cluster, described by the $G_0$ propagator, which contains a form factor related to the wavefunction of the $D^*_{s0}(2317)$ molecule \cite{Aceti}. 


Isolating the $T^{FCA}$ matrix we get:
\begin{eqnarray}\label{tdkbar}
T^{\rm FCA}_{ij}(s)=\sum\limits_{l=1}^{6}\Big[\,1-
\tilde{V}^{\rm FCA}(s)\,G_0(s)\,\Big]^{-1}_{il}\,V_{lj}^{\rm FCA}(s)\, .
\end{eqnarray}

Finally, to obtain the three-body amplitude $DK\bar{K} \to DK\bar{K}$ we need to consider that the $DK$ cluster generates the $D^*_{s0}(2317)$ in isospin 0. In order to do that we project the $T^{FCA}$ matrix in the state $|DK(I=0)\,\rangle = (1/\sqrt{2})\,|\,D^+K^0+D^0K^+\,\rangle$, obtaining a sum over the channels (1), (2), (4), (5).

One last consideration to be made is that while $T^{\rm FCA}(s)$ is written as function of the three-body center-of-mass energy $\sqrt{s}$, the $t_{i}(s_i)$ amplitudes in $V^{\rm FCA}$ and $\tilde{V}^{\rm FCA}$ are written in terms of the two-body center-of-mass energy $\sqrt{s_i}$, where $i$ stands for $D\bar{K}$ or $K\bar{K}$. To do that we use two sets of transformations to obtain $\sqrt{s_i}$ in terms of $\sqrt{s}$. 
The one we call ``Prescription I'' is a standard transformation from three-body kinematics, common in the literature, while 
``Prescription II'' is obtained assuming that the kinetic energy in the $DK$ cluster is of the order of its binding energy \cite{DKK}.

\section{Results} 

\begin{figure}[h!]
  \subfloat[]{%
    \includegraphics[width=0.47\textwidth]{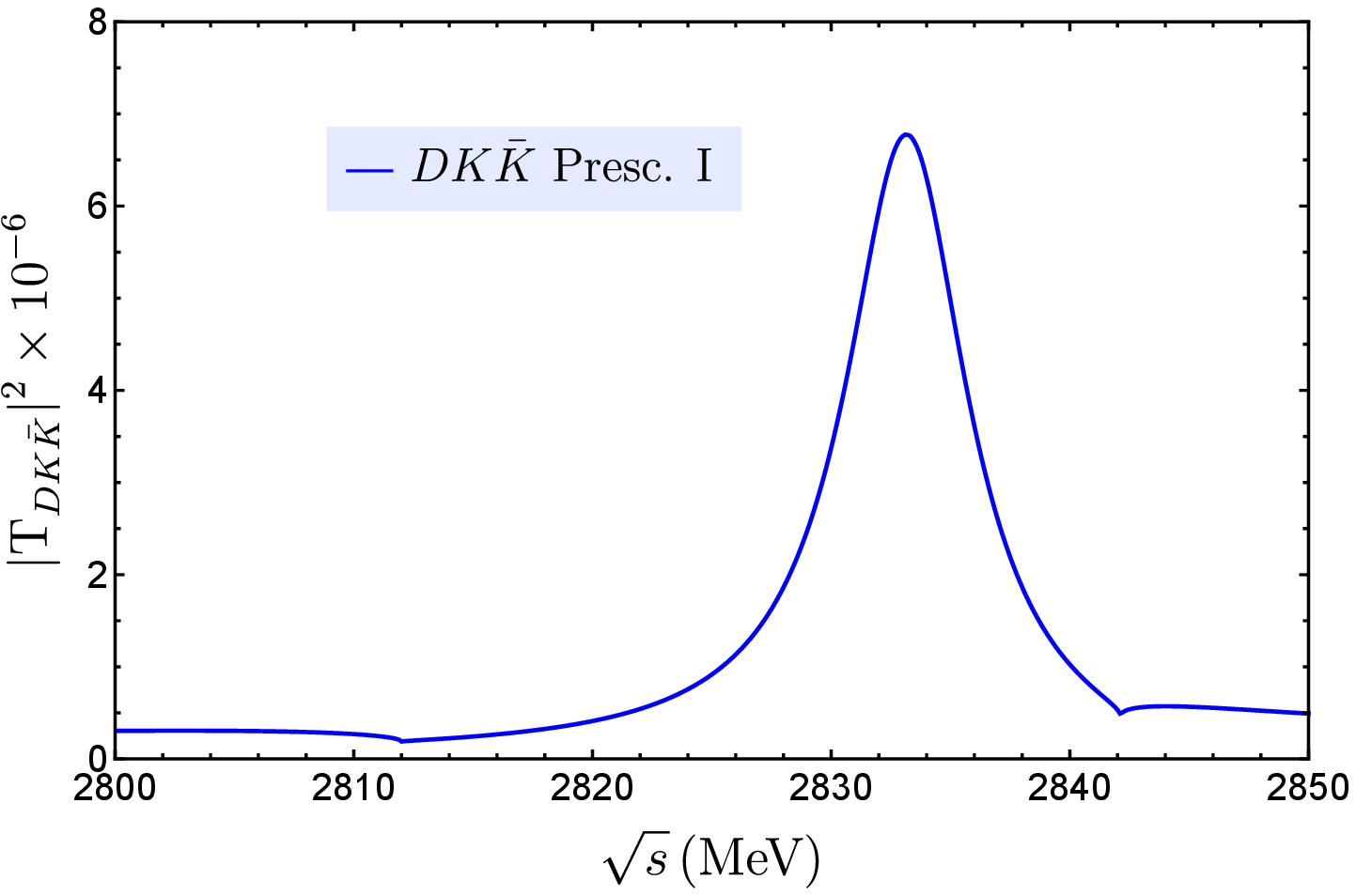} \label{fig:T2DKKbarP1}
  }
  \quad
  \subfloat[]{%
    \includegraphics[width=0.47\textwidth]{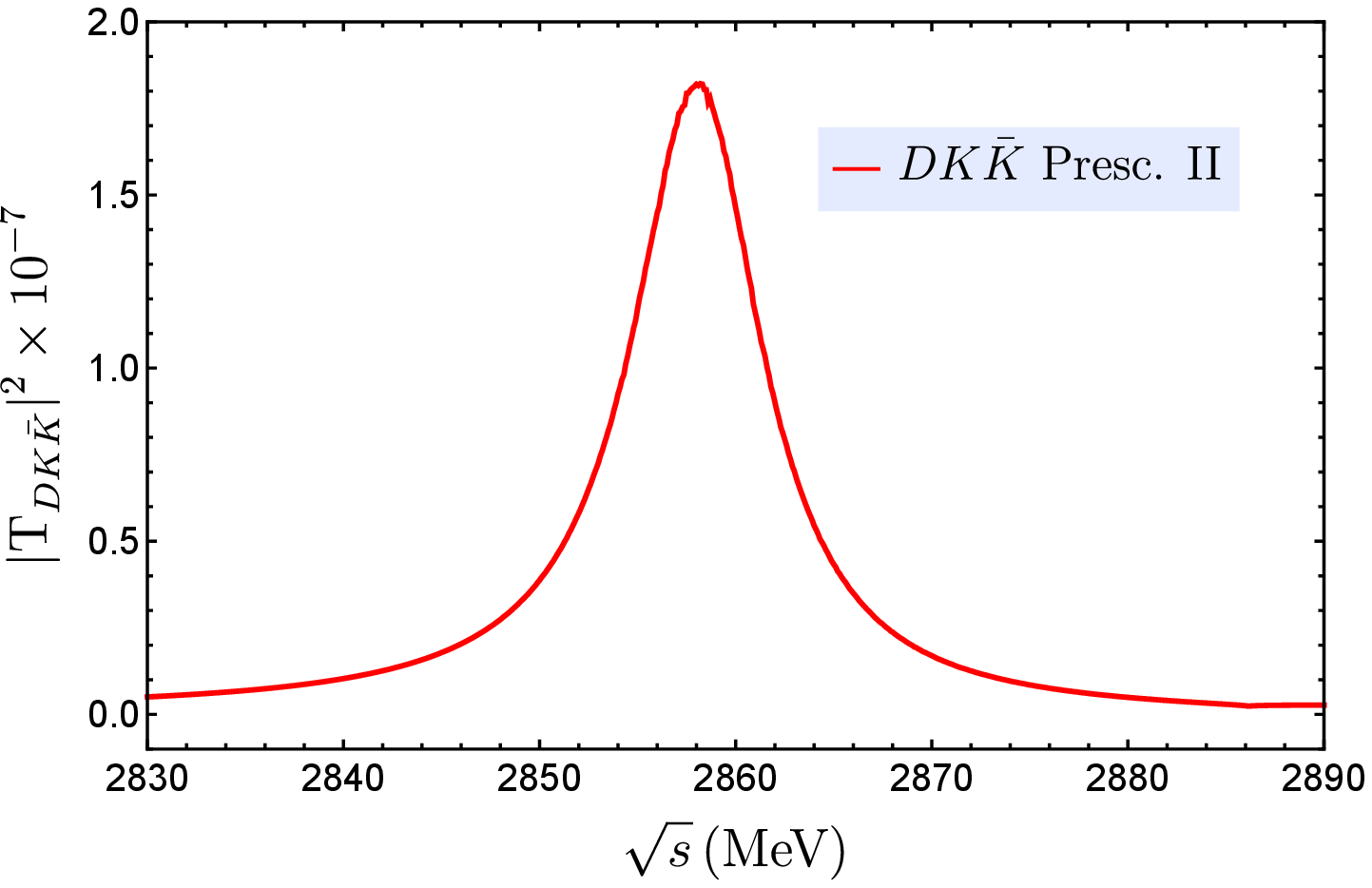} \label{fig:T2DKKbarP2}
  }
  \caption{Results for the total $D K \bar K$ amplitude squared
 using prescriptions I (left) and II (right).\label{DKKbar}}
\end{figure}

In Fig.~\ref{DKKbar} we show the results of the amplitude squared for the $DK\bar{K}$ system with the two prescriptions. We see that a narrow peak develops in both cases, indicating the presence of a state. We notice that this state is above the $D_{s0}^*(2317) \bar{K}$ threshold ($\sim2812$ MeV), but somewhat close to the threshold of $D \, f_0(980)/a_0(980)$, which would be around $2850$ MeV. Motivated by this fact and the results of Ref.~\cite{Df0} $-$ where a molecular $D \, f_0(980)$ state was found using both QCD Sum Rules and the solution of Faddeev equations without the Fixed Center Approximation $-$ we investigate the origin of this state by looking to the $K\bar{K}$ interaction, since the $D\bar{K}$ amplitude does not develop any enhancement in the energy range considered.

First we should note that the $K\bar{K} \to K\bar{K}$ scattering generates the $a_0(980)$ in isospin 1 (which also couples strongly to $\pi\eta$ in our approach using coupled channels), while the $f_0(980)$ is generated in isospin 0 (which also couples to $\pi\pi$).
\begin{figure}[h!]
  \subfloat[$K\bar{K}$ amplitude in $I=1$; couples to $a_0(980)$.]{%
    \includegraphics[width=0.47\textwidth]{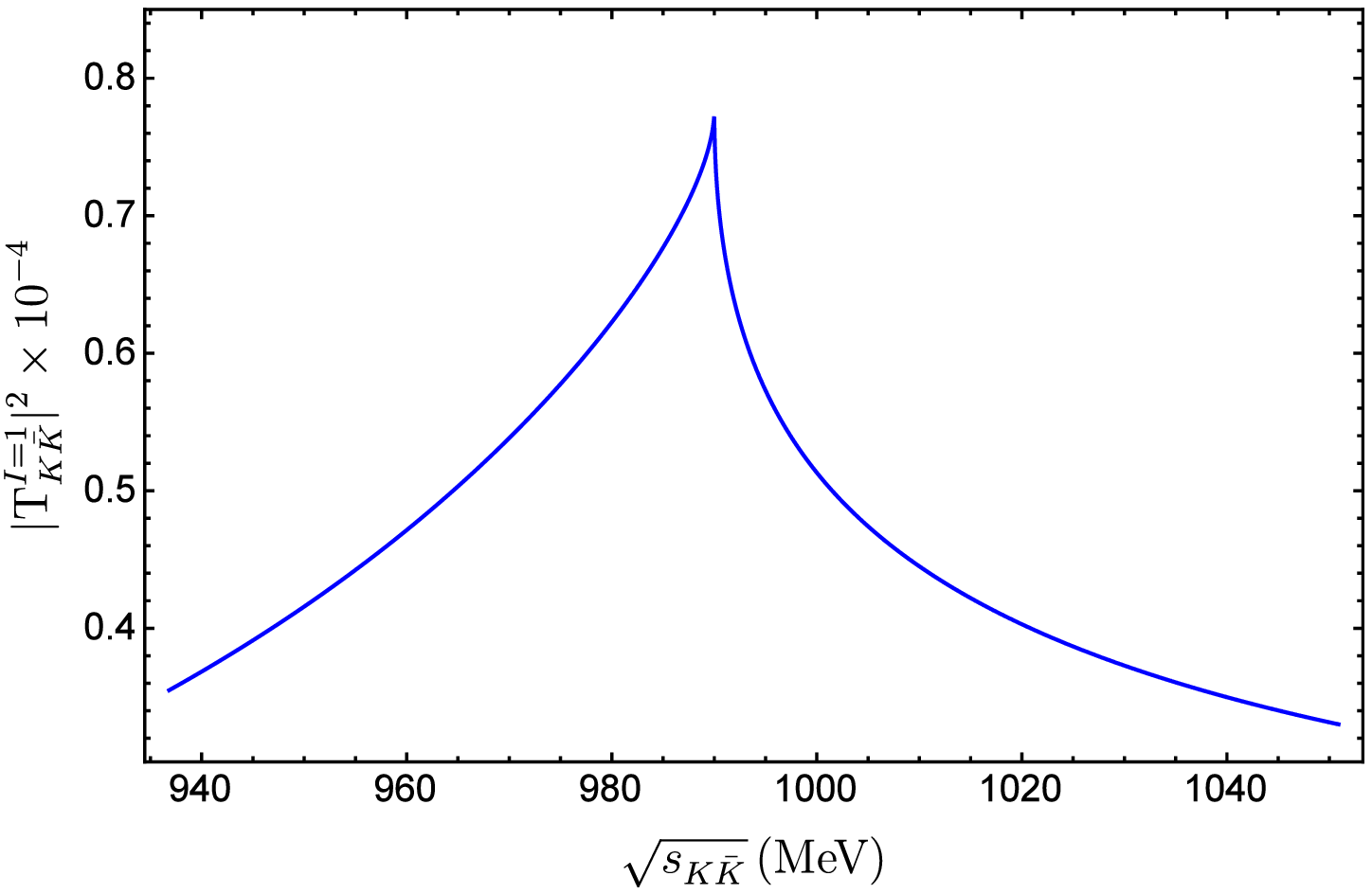} 
  }
  \quad
  \subfloat[$K\bar{K}$ amplitude in $I=0$; couples to $f_0(980)$.]{%
    \includegraphics[width=0.47\textwidth]{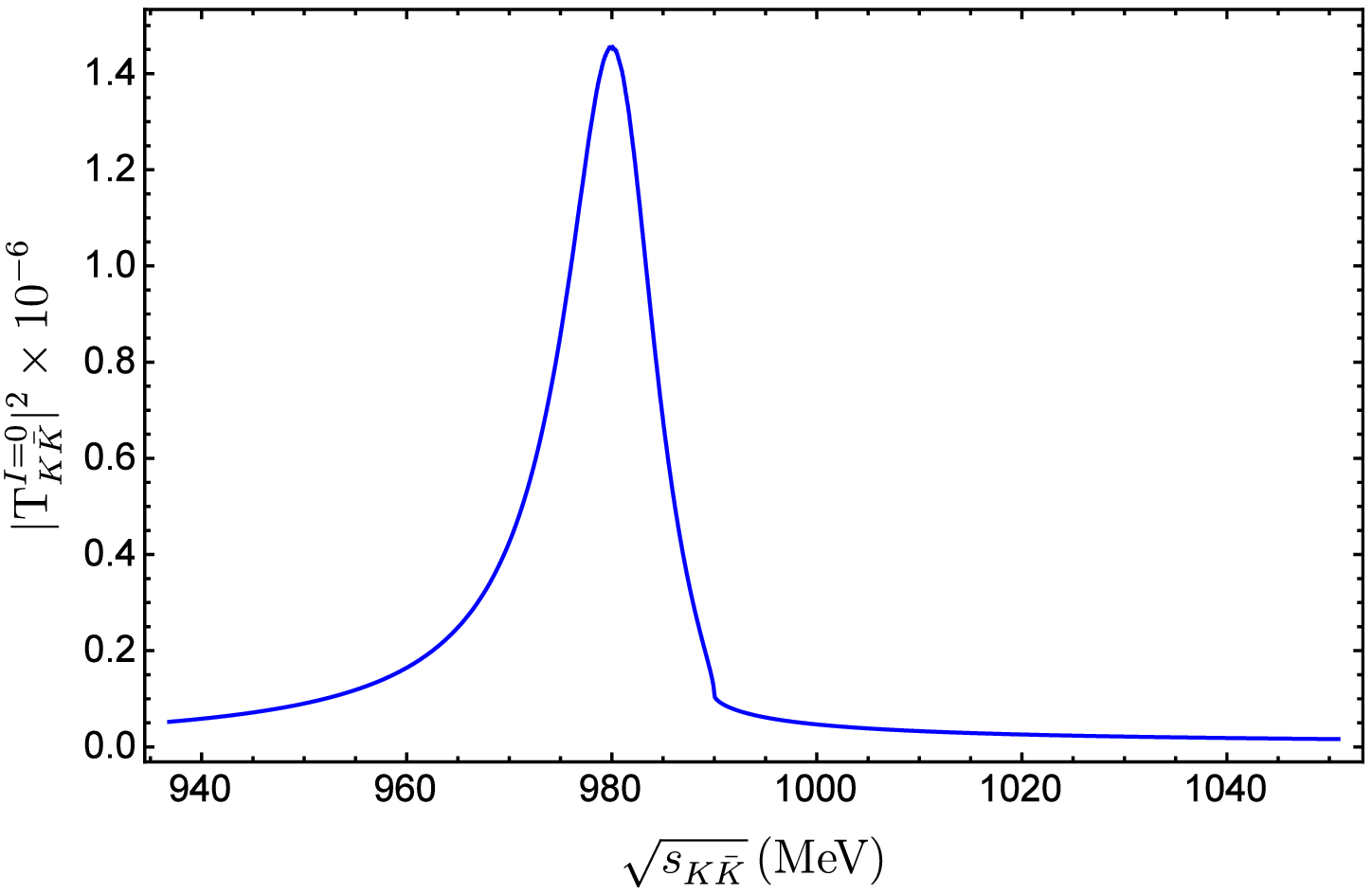} 
  }
  \caption{Comparison between $K\bar{K}$ amplitude squared in isospin 1 and 0. \label{KKbar}}
\end{figure}
In Fig.~\ref{KKbar} we compare $|T_{K\bar{K} \to K\bar{K}}|^2$, and we see that around the resonances peak the intensity of the isospin 0 amplitude is about 2 orders of magnitude stronger than in isospin 1. This is expected since the $\pi\eta$ contribution to the $a_0(980)$ is stronger than $K\bar{K}$, while the latter is the dominant one for the $f_0(980)$, in comparison to $\pi\pi$ (the $\sigma$ meson, $f_0(500)$, is dominated by $\pi\pi$ in $I=0$).

With that in mind, we write the $K\bar{K}$ amplitudes in isospin basis and switch off one at a time to see what happens with the three-body system. We find that if we remove the isospin 1 contribution the peak in the three-body amplitude remains, only shifting to the right since it becomes less bound, while if we remove the $f_0(980)$ contribution from isospin 0 the peak disappears and the intensity of $|T_{DK\bar{K}}|^2$ is reduced by about two orders of magnitude, as shown in Table \ref{Results}.


\begin{table}[h!]
\caption{\label{Results} Comparison between position and intensity of the
peaks found in the $D K \bar K$ amplitude.}
\begin{center}
\begin{tabular}{| c | c | c | c | c |}
\hline
\rule[-1ex]{0pt}{2.5ex}  & \multicolumn{4}{ c |}{ \small Prescription I      ~~                    Prescription II} \\
\hline
\rule[-1ex]{0pt}{2.5ex}  & \small $\sqrt{s}$ & \small $|T|^2$ & \small $\sqrt{s}$ & \small $|T|^2$ \\
\hline
\rule[-1ex]{0pt}{2.5ex} \small Total & \small $2833$ & \small $6.8\,\times 10^6$ & \small $2858$ & \small $1.8\,\times\,10^7$ \\
\hline
\rule[-1ex]{0pt}{2.5ex} \small $I=1$ only & \small $2842$ & \small $7.7\,\times\, 10^4$ & \small $2886$ & \small $7.8\,\times\,10^4$ \\
\hline
\end{tabular}

\end{center}
\end{table}

In view of this observation we conclude that the isospin 0 contribution is essential to generate the three-body state, and the proximity to the $D\,f_0(980)$ threshold suggests that the $K\bar{K}$ clusters around the $f_0(980)$, then this state would be mostly made of $D\,f_0(980)$, with small contributions of $D\,a_0(980)$ and $D_{s0}^*(2317) \bar{K}$.
This feature, as well as the peak position, is in fair agreement with the results of Ref.~\cite{Df0}, where
it was found $M_{Df_0}=(2926 \pm 237)$ MeV with QCD Sum Rules and $M_{Df_0} = 2890$ MeV with the solution of the Faddeev equations without FCA.\\

\begin{figure}[h!]
 \centering
 \includegraphics[width=0.6\textwidth]{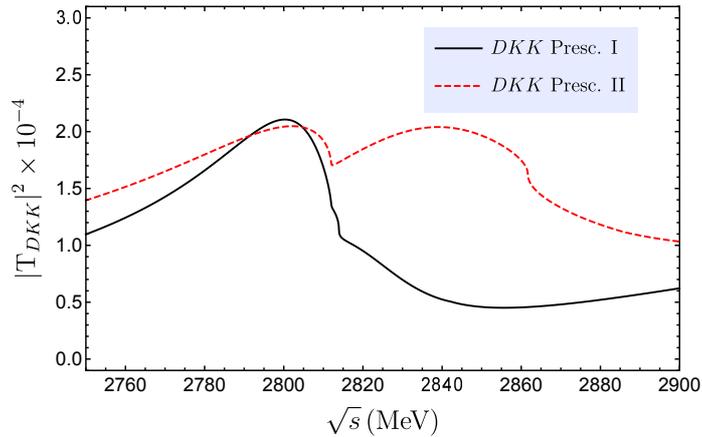}
 \caption{Results for the total $D K K$ amplitude squared using prescriptions I and II.}
 \label{DKK}
\end{figure}

On the other hand, for the $DKK$ system we do not find a clear evidence of a three-body state. We see in Fig.~\ref{DKK} that broad and irregular structures develop. Both amplitudes decrease around 2812 MeV, which corresponds to the $D_{s0}^*(2317) K$ threshold, and  with ``Prescripiton I'' there is an enhancement below this threshold. However, it seems that the $KK$ repulsion is sizable and even though the $DK$ interaction is attractive, it is not enough to bind the three mesons together. Further investigation of the $DKK$ system with the methods of Ref.~\cite{Df0} or other approaches would be interesting to confirm this result.

\section*{Conclusions}

In this work \cite{DKK} we have used the Fixed Center Approximation (FCA) to Faddeev equations, including charge exchange diagrams, to study the $DK\bar{K}$ and $DKK$ three-body systems. In both cases the $D^*_{s0}(2317)$ is taken as the heavy cluster, generated from the $DK$ interaction in isospin 0, and another $\bar{K}$ (or $K$) is added, which is allowed to undergo multiple scattering with the cluster components.
Uncertainties were estimated with the use of two different prescriptions to obtain the energy in the two-body frame as a function of the total energy of the system.

According to our results, the $DK\bar{K}$ system generates a three-body state with quantum numbers $I(J^P)=1/2(0^-)$, and mass about $2833-2858$ MeV, dominated by a $D f_0(980)$ component, where the $f_0(980)$ is implicitly contained in the $K\bar{K}$ interaction in isospin 0. These results are compatible with other methods, one using QCD Sum Rules and another with the full Faddeev equations. Since the $f_0(980)$ also couples to $\pi\pi$, this state could be seen in the $\pi\,\pi\, D$ invariant mass distribution.

The $DK K$ system, however, does not seem to bind. This is mostly due to the $KK$ repulsion, which seems to be of the same magnitude of the attractive $DK$ interaction. Further investigation of the $DKK$ system with different methods would be interesting and welcome.

\section*{Acknowledgments}

V.~R.~Debastiani wishes to acknowledge the support from the
Programa Santiago Grisolia of Generalitat Valenciana (Exp. GRISOLIA/2015/005).
J.~M.~Dias would like to thank the Brazilian funding agency FAPESP for the financial support.
This work is also partly supported by the Spanish Ministerio de Economia
y Competitividad and European FEDER funds under the contract number
FIS2014-57026-REDT, FIS2014-51948-C2-1-P, and FIS2014-51948-C2-2-P, and
the Generalitat Valenciana in the program Prometeo II-2014/068.

\end{document}